\documentclass[jkps,preprint,fleqn,showkeys]{revtex4}
\usepackage[pdftex]{graphicx}
\usepackage{amssymb}
\usepackage{amsmath}
\usepackage{bm}
\usepackage{xcolor}

\begin{document}

\title[]{Pulse shape discrimination in an organic scintillation phoswich detector using machine learning techniques}

\author{Yujin~Lee}
\author{Jinyoung~Kim}
\author{Byoung-cheol~Koh}
\author{Chang~Hyon~Ha}
\affiliation{Department of Physics, Chung-Ang University, Seoul 06974, Republic of Korea}
\author{Young~Soo~Yoon}
\affiliation{Korea Research Institute of Standards and Science, Daejeon 34113, Republic of Korea}

\begin{abstract}
We developed machine learning algorithms for distinguishing scintillation signals from a plastic-liquid coupled detector known as a phoswich.
The challenge lies in discriminating signals from organic scintillators with similar shapes and short decay times.
Using a single-readout phoswich detector, we successfully identified $\gamma$ radiation signals from two scintillating components.
Our Boosted Decision Tree algorithm demonstrated a maximum discrimination power of $3.02 \pm 0.85$ standard deviation in the 950 keV region,
providing an efficient solution for self-shielding and enhancing radiation detection capabilities.
\end{abstract}

\maketitle

\section{Introduction}
\label{sec:intro}
Organic scintillators are indispensable tools across diverse technological and scientific realms, from environmental monitoring to the investigation of rare nuclear events~\cite{Beaulieu_2016,NATTRESS20171, Kharzheev2015}. Their light-emitting property, activated when constituent molecules undergo deexcitation from ionizing radiations such as alphas, betas, gamma-induced electrons, neutron-induced protons, and cosmic-ray muons makes them vital in radiation detection, medical imaging, and nuclear physics research~\cite{birks,knoll}. Particularly, they are crucial elements in the quest to detect particles like dark matter and neutrinos, demanding highly sensitive detectors for the detection of ultra-low levels of radiation~\cite{Shirai_2017,IANNI2011405}. The appeal of organic scintillators lies in their fast decay time, ease of fabrication, and scalability, distinguishing them as a preferred choice in comparison to other scintillator options~\cite{knoll}

Dark matter particles and neutrinos, ubiquitous yet weakly interacting, pose significant challenges in measurement due to their elusive nature and poorly understood physical properties. Therefore, large-scale experiments are essential to study these particles comprehensively. Organic scintillators play a crucial role in rare decay experiments, enhancing detector capabilities through improved positioning of particle interactions. For instance, in dark matter direct detection experiments, precise positioning aids in background rejection by distinguishing radioactivity in the surrounding environment from that of the main target material~\cite{IANNI2011405}. Segmentation concepts, particularly beneficial in short baseline neutrino experiments, exploit variable oscillation baselines within a single experiment~\cite{PROSPECT:2018dnc,Abreu_2021}.

This study introduces a novel single-readout detector called a phoswich (phosphor sandwich)~\cite{phosphor} where plastic scintillator (PS) serves as the inner target, and liquid scintillator (LS) acts as the outer guard. Traditional light sensors such as Photomultiplier Tubes (PMTs), containing a significant amount of natural U/Th/K radioactivities, tend to elevate the background level, particularly when placed in close proximity to the target material. The phoswich target-guard approach can improve the target's background contamination by physically distancing PMTs and identifying interactions detected in the guard material, which also serves as a light guide.
In addition to pinpointing interactions, the identification of particles such as gammas, alphas, and neutrons with signal shape analyses could enhance detector sensitivity with a reduced number of PMTs per target volume.
Despite the challenge posed by the similar decay times of a few nanoseconds for the two organic scintillators within the phoswich setup, this study utilizes machine learning techniques to effectively discriminate gamma signals between the two scintillators,
marking an important first step towards a large-scale position-sensitive low-background detector concept.

\section{Materials and Methods}

\subsection{Experimental setup}

In our setup, a phoswich configuration is implemented, wherein two scintillators with distinct pulse shapes are optically coupled and share a common readout system through PMTs. The implementation of pulse shape discrimination (PSD) serves to segregate signals originating from each scintillator, facilitating the identification of radiation interaction types and, consequently, their source locations.

The primary objective in designing the detector for this analysis is to efficiently collect scintillation photons from the two scintillators. To achieve this, we utilize a plastic scintillator as the inner material and a liquid scintillator as the guard material.
The detector configuration comprises a reflective polytetrafluoroethylene (PTFE) cylinder which houses the inner plastic scintillator, encapsulated by a thin polymethylmethacrylate (PMMA) case positioned in the middle of the main cylinder. Importantly, the PMT photocathodes are in direct contact with the liquid scintillator, eliminating the need for any intermediate optical interfaces.

\begin{figure}[!htb]
  \begin{center}
      \includegraphics[width=0.48\textwidth]{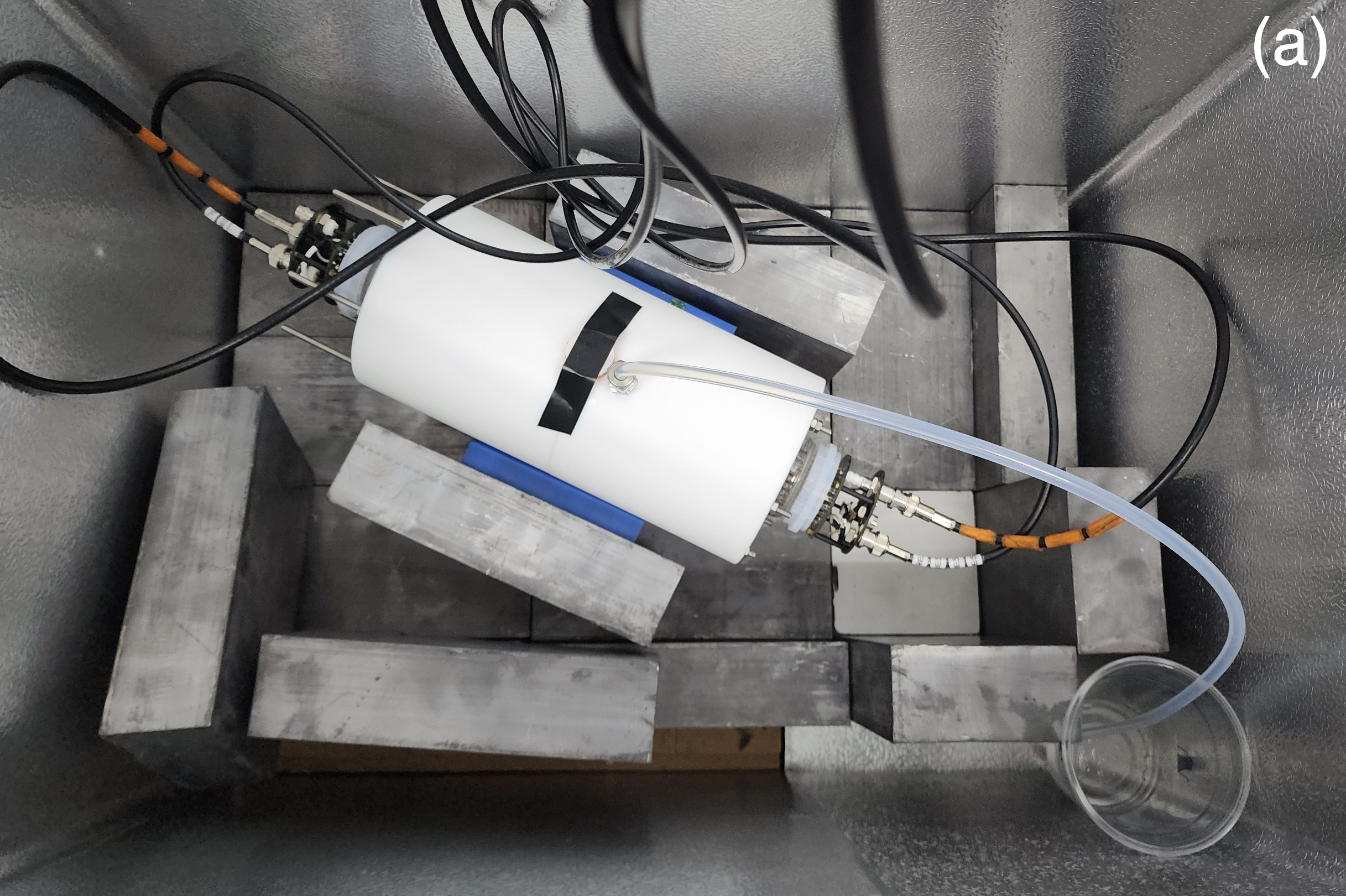}
      \includegraphics[width=0.47\textwidth]{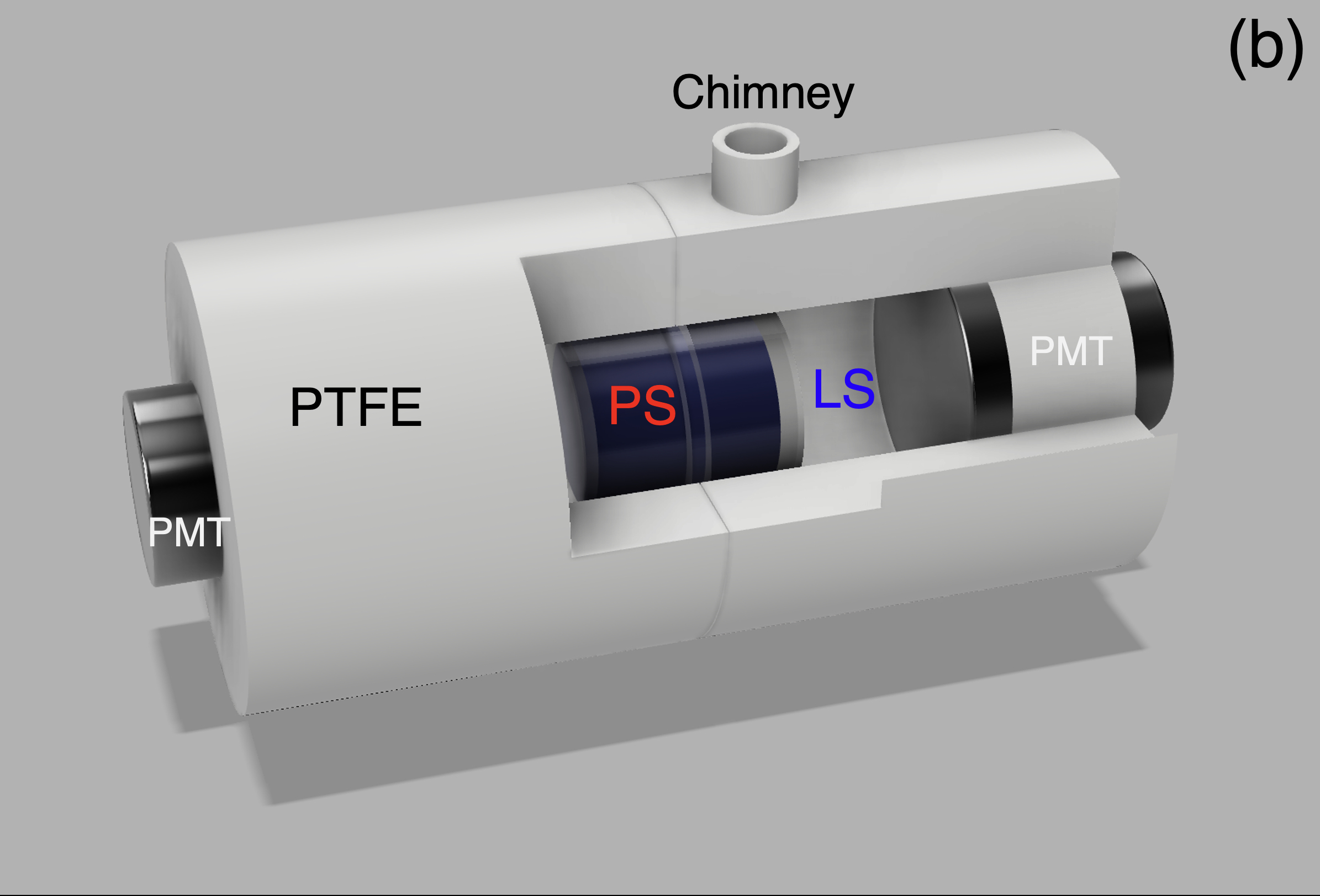}
  \end{center}
  \caption{Phoswich detector. \textbf{(a)} The photo of the setup for the detector and the dark box, \textbf{(b)} The drawing of the detector with internal area exposed. The 3-inch PMTs are attached to both sides of the detector to measure the photons emitted from the guard liquid and inner plastic scintillator.
  LS is introduced into and removed from the chimney, which also prevents overflow caused by thermal expansion.
  }
  \label{phoswich}
\end{figure}

The PS~\footnote{Eljen Technology EJ-200} component is precision-machined into a cylindrical shape with dimensions of 45.7~mm in diameter and 49~mm in length. The surfaces of the PS cylinder undergo a polishing process using lapping films to achieve high transparency. To protect PS from chemically reacting with LS, a cylindrical container is fabricated with a diameter of 50.3~mm, a length of 59~mm, and a thickness of 2~mm. Two 1~mm-thick lids made of PMMA are prepared for sealing. The PS cylinder is placed within the PMMA container, and its surfaces are optically coupled by filling small gaps with silicon grease~\footnote{EJ-550 is used}. Subsequently, the container is sealed by attaching the lids using optical cement and adhesives.

For the main container, two cylindrical tubes made of 22~mm-thick PTFE are prepared, each with an inner diameter of 76~mm and a length of 120~mm. One end of these tubes features a 53~mm hole where the PMT rear glass part is fitted and sealed with an O-ring, while the other end is left open to align with the second half of the assembled structure. When the PMMA cylinder is centered within the PTFE container, the outer PTFE components are connected using an O-ring and six long connecting bolts through PTFE flanges. A small PMMA support piece stabilizes the inner PS assembly at the center of the PTFE tube. The inner surface of the PTFE container serves as a diffusive light reflector. The container is filled with LS through the hole under the chimney.

The properties of the LS counter have been extensively examined in experiments requiring detectors with substantial volume or irregular shapes~\cite{Wen_2017}. Linear alkylbenzene (LAB), chosen for its non-toxic and cost-effective attributes, is used as the LS solvent. For initial fluorescence, a concentration of 3~g/L of 2,5-diphenyloxazole is dissolved in the solvent, and for a wavelength shift, 30~mg/L of p-bis(o-methylstyryl) benzene is added~\cite{Adhikari:2017esn}.

The detector assembly features two three-inch PMTs manufactured by Hamamatsu (R12669SEL). These PMTs show a typical gain of $1.0\times10^6$, a spectral response range of 300 to 650~nm, and a peak wavelength of 420~nm. The entire assembly is placed within a dark box with lead shielding averaging 50~mm in thickness. Figure~\ref{phoswich} illustrates the completed experimental setup and a schematic drawing of the detector.

\subsection{Readout system and data collection}

The data acquisition process involves passing signals through an analog-to-digital converter (FADC), specifically employing a 500~Megasamples/s FADC with a 2.5~V, 12-bit dynamic range. The custom Data Acquisition (DAQ) system, provided by the local company Notice Korea, has a proven track record in various experiments~\cite{Adhikari:2018fpo}.

For event construction, an 8000~ns readout window is established when both PMTs register coincidence pulses within 200~ns, each requiring a discrimination threshold of 10~ADC (where 1~ADC corresponds to approximately 0.6 mV). Utilizing delay settings, the positive pulse starting position is set near 2400~ns, and pedestals are subtracted on an event-by-event basis by averaging the first 500~ns region where no signal is present. High voltages are carefully adjusted such that the PMT signal pulses with less than about 2~MeV energy are not saturated. Synchronization of the two PMT gains is achieved through fine-tuning each high voltage using the Compton edges of the $^{60}\rm Co$ gamma peaks. The raw data in binary format are subsequently converted into ROOT format for post-processing~\cite{root}.

Regular gamma calibration data are obtained using $^{60}\rm Co$ (1173.2 and 1332.5~keV) and $^{137}\rm Cs$ (661.7~keV) sources.
This calibration ensures the accuracy and reliability of the collected data throughout the experimental duration.

\subsection{Pulse Shapes}

Typical organic scintillator pulses are fast having a decay time less than 5~ns and rising time of about 2~ns.
The decay time for PS is 2.1~ns~\cite{EJPS} while that of LAB-based LS is 4.0~ns~\cite{ZHONG2008300}. 
With a convolution of PMT transit time of 60~ns (9.5~ns rise time and 13~ns spread), the shape discrimination of output signal waveforms from two scintillators are non-trivial problem, requiring a careful analysis.

\begin{figure}[!htb]
  \begin{center} 
      \includegraphics[width=0.65\textwidth]{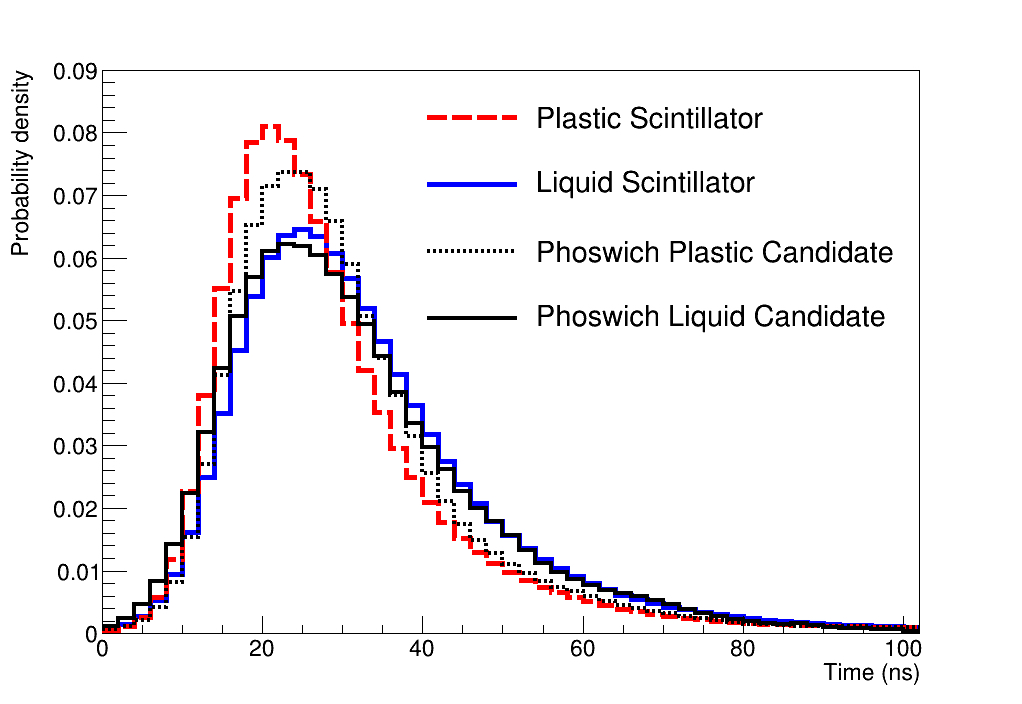}
  \end{center}
  \caption{Pulse shape distributions of each scintillator. 
    The red dashed line indicates the PS average waveform as a probability density distribution while the blue solid line is for LS.
    The black dotted and solid lines are PS-candidate and LS-candidate events, respectively from the phoswich data. The PS waveforms show a faster rise and decay in time than LS.}
  \label{pulse}
\end{figure}

Figure~\ref{pulse} shows the average waveform distributions of the PS (red dashed line) and the LS (blue solid line), normalized by the area and for the same energy region. These separate measurements are done by two independent detector assemblies with a $^{60}$Co source where the target materials are directly coupled with two PMT photocathods. The start time ($\rm t_0$) of the waveform is defined as the time at which a pulse crosses 6-ADC threshold. The distributions show that the leading edge and trailing edge of the PS waveform are relatively steeper than those of the LS waveform. Therefore, we explored this shape characteristics by developing various discriminating variables.

Additionally, we examined representative candidate events from the phoswich data. One candidate event (depicted by the black solid line) originates from the liquid segment of the phoswich, while the other (depicted by the black dotted line) is derived from the plastic component, as superimposed in Figure~\ref{pulse}. The signal shape of the phoswich-PS part exhibits noticeable degradation, especially at the leading edge, when compared to the PS-only measurement. This degradation suggests a potential scenario where early scintillation photons might have encountered scattering and/or reabsorption by the guard LS before reaching a PMT. Nonetheless, the shape information of the phoswich detector remains intact, and we delved into the subtle yet conceivable shape differences by formulating various parameters.

\subsection{Discrimination Variables}
The objective of parameter development is to effectively discern the pulse shape differences between two scintillators utilizing measurements from the $^{60}$Co source. Given that the primary characteristic of the waveforms is the time-amplitude structure, our focus is on fractions of amplitudes in different parts of the waveform, average times, their variances, and the degree of template matches.

\begin{figure}[!htb]
  \begin{center}
      \includegraphics[width=0.6\textwidth]{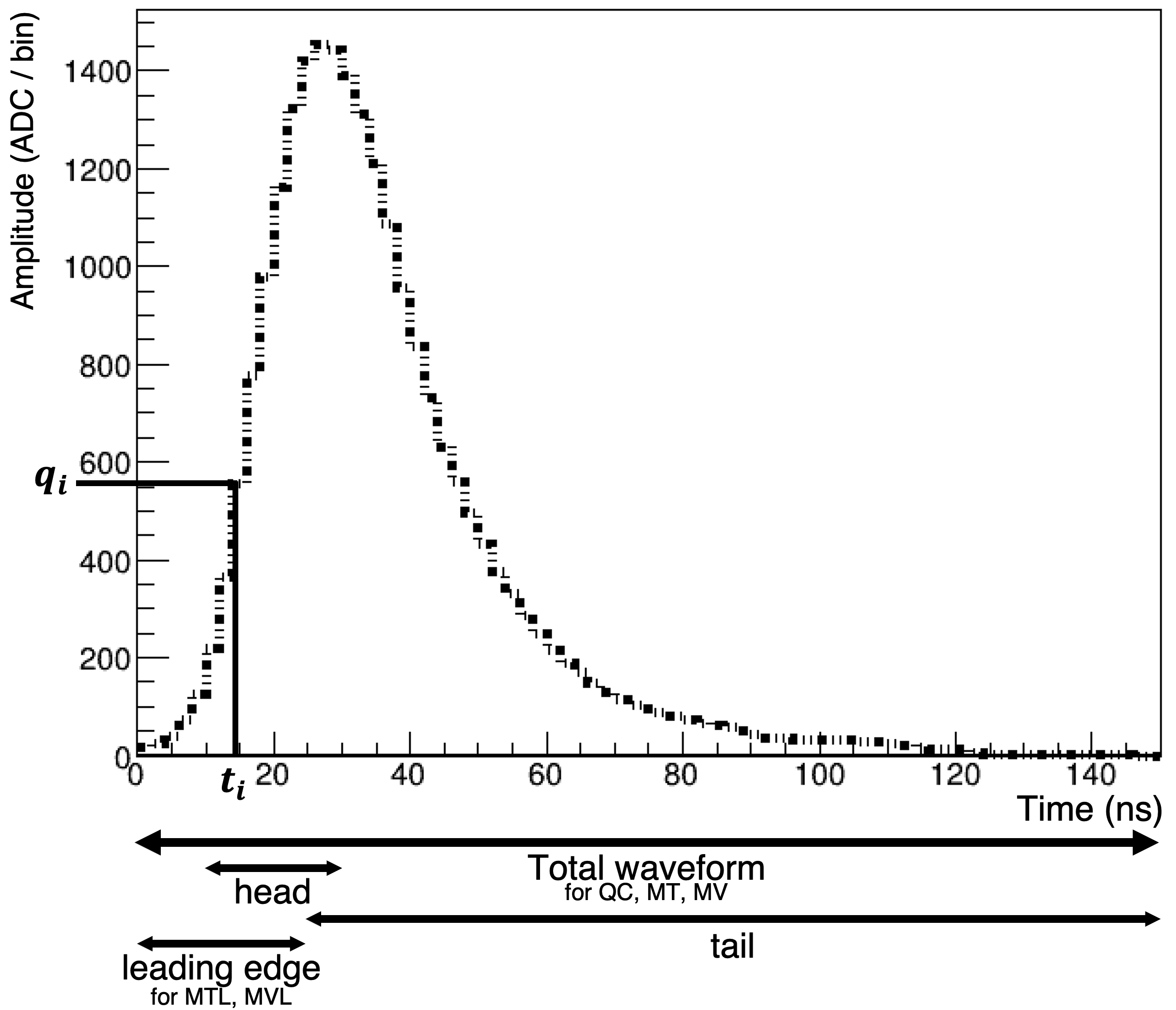}
  \end{center}
  \caption{Variable calculation methods for a waveform.
    An example LS waveform and the diagram for representing the parameters and the time range of the variables are displayed.}
  \label{var_LS}
\end{figure}

\begin{figure}[!htb]
  \begin{center}
      \includegraphics[width=0.95\textwidth]{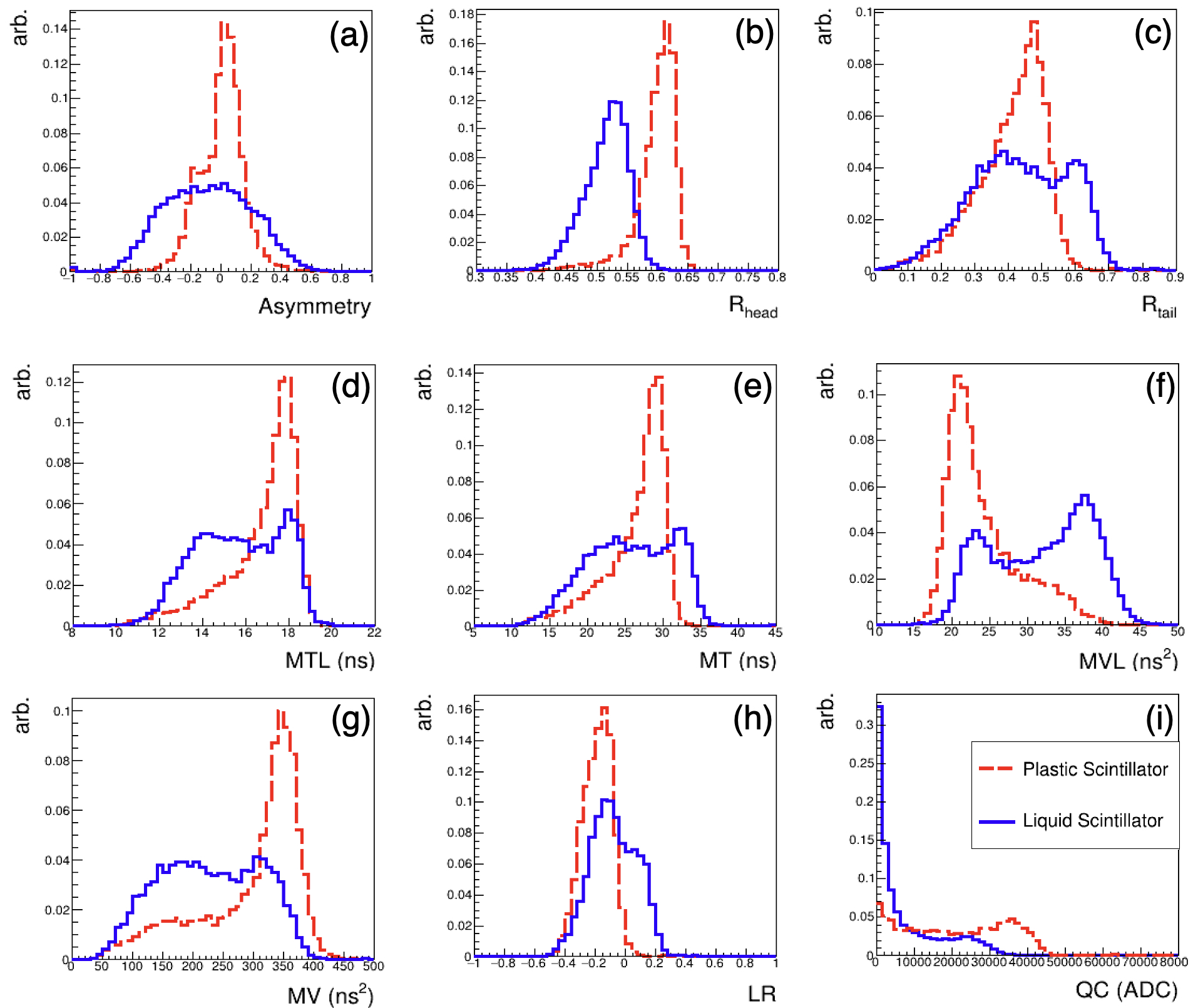}
  \end{center}
  \caption{Discrimination variables. These variables are labeled as \textbf{(a)} Asymmetry, \textbf{(b)} $\rm R_{head}$, \textbf{(c)} $\rm R_{tail}$, \textbf{(d)} MTL, \textbf{(e)} MT, \textbf{(f)}, MVL \textbf{(g)} MV, \textbf{(h)} LR, and \textbf{(i)} QC. Please see the text for the variable descriptions.}
  \label{var}
\end{figure}

We use the balance of the deposited charge from two PMTs, Charge Asymmetry (Eq.~\ref{eq:asym} also shown in \textbf{(a)} of Fig.~\ref{var})~\cite{Adhikari:2017esn}, 
the ratio $\rm R_{head}$ of the integrated charge in the 10~ns to 30~ns time window (head)
to the total collected charge in the first 150~ns time span (Eq.~\ref{eq:qhead} also shown in (b) of Fig.~\ref{var})~\cite{2021102581}, 
the ratio $\rm R_{tail}$ of the integrated charge in the 25~ns to 150~ns time window (tail)
to the total collected charge in the first 150~ns time span (Eq.~\ref{eq:qtail} also shown in (c) of Fig.~\ref{var})~\cite{Ko_2016}, 
the charge–weighted mean time MT of pulses within first 25~ns and 150~ns
(Eq.~\ref{eq:mt} also shown in \textbf{(d)} and \textbf{(e)} of Fig.~\ref{var})~\cite{Gerbier:1998dm}, 
the variance of the charge–weighted mean time MV
(Eq.~\ref{eq:mv} also shown in \textbf{(f)} and \textbf{(g)} of Fig.~\ref{var})~\cite{govinda}, 
the likelihood ratio parameter LR using PS and LS waveform templates in first 150~ns
(Eq.~\ref{eq:lr} also shown in \textbf{(h)} of Fig.~\ref{var})~\cite{2021102581}, 
and the total charge QC (Eq.~\ref{eq:qc} also shown in \textbf{(i)} of Fig.~\ref{var}). The LS waveform with diagrams for understanding these variables is shown in Fig.~\ref{var_LS}. The definitions of each variable used in the PSD algorithms are following,
\begin{eqnarray}
  Charge~~Asymmetry = (Q_1 - Q_2)/(Q_1 + Q_2) \label{eq:asym}\\
  R_{head} = \sum^{30\,ns}_{10\,ns}{q_i}/\sum^{150\,ns}_{0\,ns}{q_i}   \label{eq:qhead}\\
  R_{tail} = \sum^{150\,ns}_{25\,ns}{q_i}/\sum^{150\,ns}_{0\,ns}{q_i}   \label{eq:qtail} \\
  MTL = \sum^{25\,ns}_{0\,ns}{q_it_i }/\sum^{25\,ns}_{0\,ns}{q_i} \quad \textrm{and} \quad MT = \sum^{150\,ns}_{0\,ns}{q_it_i }/\sum^{150\,ns}_{0\,ns}{q_i}   \label{eq:mt}\\
  MVL = \sum^{25\,ns}_{0\,ns}{q_it_i^2 }/\sum^{25\,ns}_{0\,ns}{q_i} - MTL^2 \quad \textrm{and} \quad MV = \sum^{150\,ns}_{0\,ns}{q_it_i^2 }/\sum^{150\,ns}_{0\,ns}{q_i} - MT^2   \label{eq:mv} \\
  LR = \ln L_{ps} - \ln L_{ls}, ~where~ L_{ps(ls)} = \sum_{0\,ns}^{150\,ns}(T_i-W_i+W_i\ln \frac{W_i}{T_i})   \label{eq:lr}\\
  QC = \sum^{150\,ns}_{0\,ns}{q_i}  \label{eq:qc},
\end{eqnarray}
where Q$_{1,2}$ indicates individual PMT integrated charges, and q$_i$ and t$_i$
are waveform amplitudes and times for each 2\,ns bin, respectively.
T$_{i}$ and W$_{i}$ are the heights of the i$^{th}$ time bin in the template and the data waveform, respectively.
With the exception of the Charge Asymmetry and QC parameters, an average value is used for the calculations of all other parameters, considering both PMTs.

The LR parameter is derived through the use of templates for each scintillator shown in Fig.~\ref{pulse}. After obtaining the normalized waveform for a single event, the heights of the normalized waveform and the corresponding template for each time bin are incorporated into the Poisson-based likelihood ($\rm L_{ps(ls)}$ in Eq.~\ref{eq:lr}). The template preferences for each scintillator are established, and the difference between them is employed as the LR variable. For example, in the case of an LS candidate event normalized to have the same area as the templates in Fig.~\ref{pulse}, closely resembling the LS template, the LR value is situated within the LS region, notably distant from the plastic scintillator (PS) region (plot \textbf{(h)} of Fig.~\ref{var}).

\section{Results with the Boosted Decision Tree algorithm}
\begin{figure}[!htb]
  \begin{center}
      \includegraphics[width=0.95\textwidth]{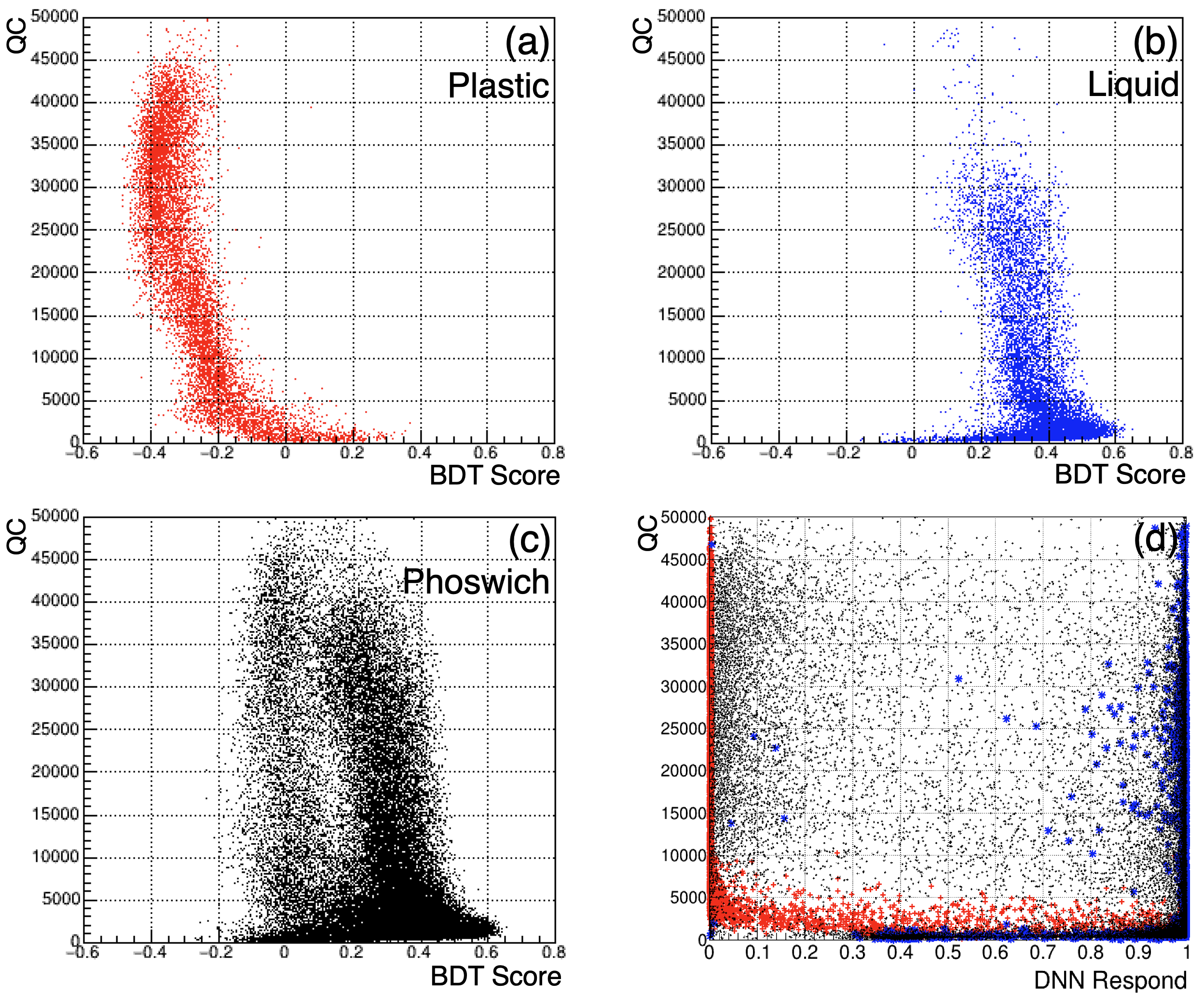}
  \end{center}
  \caption{BDT and DNN result distributions of scintillators. \textbf{(a)} BDT result of the PS training data, \textbf{(b)} BDT result of the LS training data and \textbf{(c)} BDT result of the phoswich detector. \textbf{(d)} represents the DNN results, the red cross for the PS, the blue star for the LS, and the black dot for the phoswich.
    Please note that the BDT response in the phoswich data is different from the training data especially for the PS part
    because the light from the phoswich PS part is attenuated more than that of the training setup.
  }
  \label{bdt2}
\end{figure}

Boosted Decision Trees (BDTs) were trained using above-mentioned eight variables except QC within TMVA packages~\cite{Speckmayer:2010zz}.
BDT is an ensemble learning method that combines the outputs of multiple weak learners (decision trees)
to create a stronger, more accurate model.
We use a BDT with 850 trees, each with a maximum depth of 3 levels, and AdaBoost~\cite{schapire2013explaining}
with a learning rate of beta=0.5.
A constraint on the minimum number of samples for node splitting is set to 2.5\% of the total samples.
We use a total of 9940 and 9984 events in the BDT input as PS and LS sample, respectively.
These samples are further divided in half for training and testing process. 
The resulting training shows no overtraining when evaluated with the Kolmogorov-Smirnov (KS) test between training and testing sample~\cite{kstest}. The PS sample shows the KS probability value of 9.7\% and it is 38.9\% for LS.
The MVL parameter shows the best performance out of the training.
The electron/gammas signal models for each scintillator is obtained from the $^{60}$Co calibrations of the separate measurements. The training shows a clear separation for PS and LS training samples as shown in Fig.~\ref{bdt2}.

\begin{figure}[!htb]
  \begin{center}
      \includegraphics[width=0.6\textwidth]{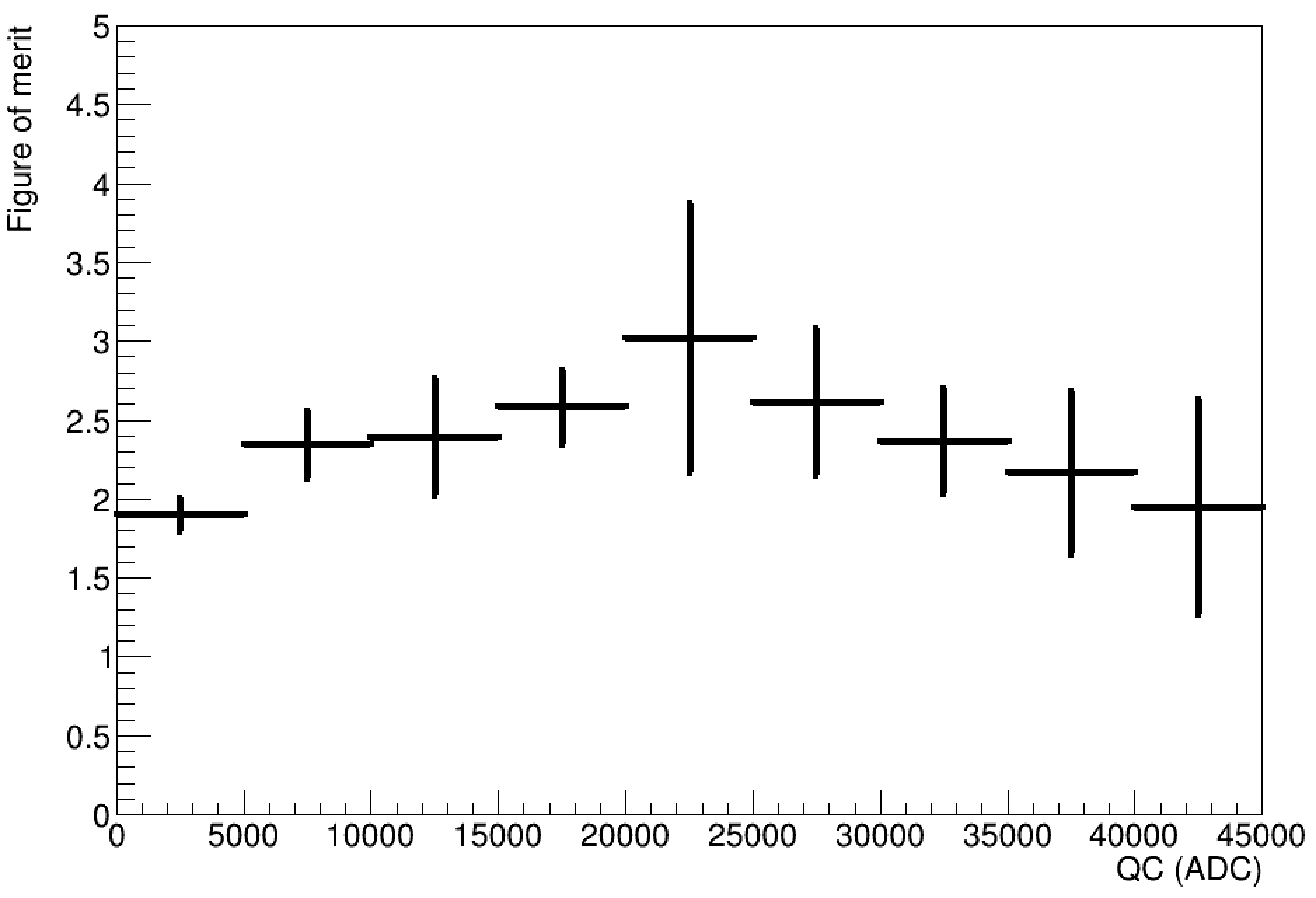}
  \end{center}
  \caption{Figure of merit as a function of QC. The maximum figure of merit is reached at QC between 20,000~ADC and 25,000~ADC.
    Note that the distribution has a slight charge dependence due to $\rm t_0$ variations in waveforms. 
  }
  \label{bdt3}
\end{figure}

\begin{figure}[!htb]
  \begin{center}
      \includegraphics[width=1.0\textwidth]{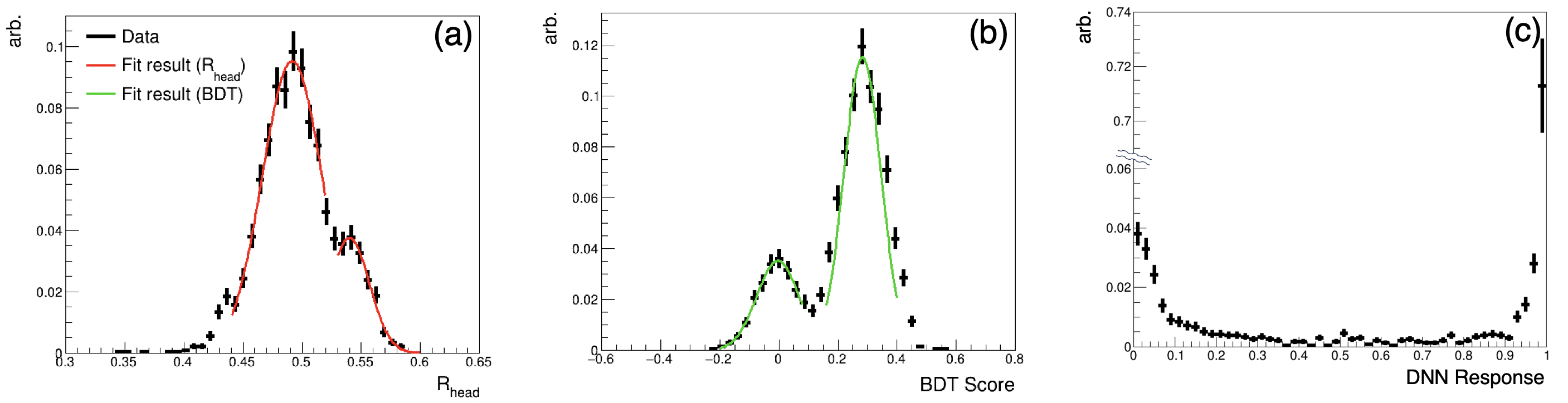}
  \end{center}
  \caption{Distribution of \textbf{(a)} $\rm R_{head}$, \textbf{(b)} BDT score, and \textbf{(c)} DNN response of the phoswich from 20,000~ADC to 25,000~ADC.
    For the $\rm R_{head}$ and BDT parameters, Gaussian fit lines are provided on the data points where smaller bumps indicate the PS part of the phoswich data.
    Note that the middle part of the y-axis of the right plot has been suppressed to increase the readability.}
  \label{bdt4}
\end{figure}

The trained machine is, then, applied to newly acquired $^{60}$Co data for the phoswich detector showing the separation between PS and LS originating signals within the phoswich detector.
To quantify the discriminating power, we calculated significance of separation using the figure of merit (FoM) as
\begin{eqnarray}
  \rm FoM = \frac{|\langle m_{LS} \rangle - \langle m_{PS} \rangle |}{\sqrt{\sigma_{LS}^2 +\sigma_{PS}^2}} \label{eq:fom},
\end{eqnarray}
where  $\rm \langle m_{LS(PS)} \rangle$ indicates the means of each distribution and $\rm \sigma_{LS(PS)}$ for widths obtained from a Gaussian fit respectively for PS and LS after selecting sections of the QC energy proxy. The FoM distribution calculated from Eq.~\ref{eq:fom} as a function of the total charge (QC) is shown in Fig.~\ref{bdt3}. 

The best separation power, FoM$=$3.02$\pm$0.85 as a standard deviation is obtained at the QC region between 20,000~ADC and 25,000~ADC which is the LS energy range equivalent between 859~keV and 1036~keV. The power is degraded as the energy is decreased due to small signals having lack of information. By comparing with a single variable FoM using popular PSD parameter $\rm R_{head}$ (FoM$=$1.59) shown in Fig.~\ref{bdt4}, BDT based method improves the separation by a factor 1.90. We checked and confirmed that the target volume ratio between PS and LS of 3.38$\pm$0.01 matches with the selected event count ratio of 3.41$\pm$0.17. The standard deviation reaches to 1 unit at the threshold of 297 keV below which pulses are not distinguishable with the current method.

We conducted tests using alternative machine learning algorithms on the same PSD data, using identical input parameters to assess algorithmic stability. In Fig.~\ref{bdt2} (d), it is evident that the PS-only and LS-only data are well-differentiated towards the extremes (0 or 1) of the Deep Neural Network response. However, the phoswich data displays a distribution where the majority of events are distinctly assigned to either the PS or the LS region, yet some events in between are not effectively separated. Moreover, Fig.~\ref{bdt4} (c) indicates that a significant proportion of events are misclassified as LS events, mismatching the target volume ratio. Consequently, the BDT method is more suitable for effectively distinguishing signals from the phoswich.

\section{Discussion}
In the phoswich detector, while the pulse shape of the guard LS closely resembles the training dataset, the pulse shape from the inner PS differs from the training data. Consequently, the BDT scores for PS are attenuated in the phoswich setup. This discrepancy arises because the training data for PS is acquired by directly attaching the PMTs to the surface of the PS material before the phoswich assembly, where LS is present between PS and PMT photocathodes. Consequently, scintillation light from PS must traverse through LS, with some photons being either reabsorbed by LS or excessively scattered and reflected within the phoswich, resulting in a loss of the original shape information.

To enhance the separation power, we anticipate that improvements in the light attenuation quality of LS would be beneficial. Therefore, the energy threshold defined at the moment would be improved. Specifically, refining the light attenuation characteristics of LS could potentially mitigate the impact of signal distortion within the phoswich setup, leading to a more accurate and reliable discrimination of scintillation signals from the inner PS. We have examined the improved LS spectra in resolutions and gains when the nitrogen bubbling is performed to remove oxygen impurities in LS.
The method was effective to improve the quality of LS for a short term.

Employing a single-readout phoswich detector, we successfully identify $\gamma$ radiation signals from two distinct scintillating components. The BDT algorithm demonstrates a discrimination power of 3.02~$\sigma$ between the two scintillators. Our exploration of signal separation using the DNN algorithm indicates that the BDT is better suited for this task. Future efforts will focus on enhancing the detector to have a possible neutron versus $\gamma$ classification in a phoswich detector.

\section*{Conflict of Interest Statement}
The authors declare that the research was conducted in the absence of any commercial or financial relationships that could be construed as a potential conflict of interest.

\section*{Author Contributions}

CHH contributed to the conception and design of the study.
YJL, JK, and BCK built the experimental setup with the datataking software tools.
YJL organized the data acquisition and performed the statistical analysis.
YSY provided the calibration facility.
YJL and CHH wrote the first draft of the manuscript.
All authors contributed to manuscript revision, read, and approved the submitted version.

\section*{Funding}
This research was supported by the Chung-Ang University Research Grants in 2022 and
by the National Research Foundation of Korea~(NRF) grant funded by the Korean government~(MSIT) (No. 2021R1A2C1013761).

\section*{Data Availability Statement}
The datasets analyzed for this study can be provided with a reasonable request to the corresponding authors.

\end{document}